\pdfoutput=1
\documentclass[runningheads]{llncs}

\usepackage{graphicx}
\usepackage{subfig}
\usepackage{multirow}
\usepackage{url}
\usepackage{comment}
\usepackage{booktabs}
\usepackage[frozencache,cachedir=.]{minted}

\newcommand{\squeezeup}{\vspace{-2.2mm}}

\begin{document}

\title{Particle-In-Cell Simulation\\ using Asynchronous Tasking}

\author{
Nicolas Guidotti\inst{1}\and Pedro Ceyrat\inst{1} \and João Barreto\inst{1} \and José Monteiro\inst{1} \and Rodrigo Rodrigues\inst{1} 
\and Ricardo Fonseca\inst{2,3}  \and Xavier Martorell\inst{4} \and Antonio J. Pe\~na\inst{4} 
}

\authorrunning{Guidotti et al.}

\institute{INESC-ID, Instituto Superior Técnico, Universidade de Lisboa \and IPFN, Instituto Superior Técnico, Universidade de Lisboa \and DCTI/ISCTE-IUL \and Barcelona Supercomputing Center (BSC)}

\maketitle

\begin{abstract}

Recently, task-based programming models have emerged as a prominent alternative among shared-memory parallel programming paradigms. Inherently asynchronous, these models provide native support for dynamic load balancing and incorporate data flow concepts to selectively synchronize the tasks. However, tasking models are yet to be widely adopted by the HPC community and their effective advantages when applied to non-trivial, real-world HPC applications are still not well comprehended. In this paper, we study the parallelization of a production electromagnetic particle-in-cell (EM-PIC) code for kinetic plasma simulations exploring different strategies using asynchronous task-based models. Our fully asynchronous implementation not only significantly outperforms a conventional,  synchronous approach but also achieves near perfect scaling for 48 cores.

\end{abstract}
\section{Introduction}
\label{sec:intro}

As the number of processing units in multicore processors increases, so does the
overhead for running parallel 
applications on these systems. Many alternative programming models have been
proposed to facilitate the software development, while achieving higher application
efficiency. Among them, task-based programming has long been hailed for its good
load-balancing features, and has reached the mainstream with its adoption on OpenMP (\texttt{task} directive).
Since task constructs
are inherently asynchronous, they have the potential to prevent synchronization points in the code. Such synchronization can cause idle periods in processors, hence reducing
performance and efficiency of HPC applications. 
Moreover, programming using tasks is
increasingly being used as a means to facilitate the development 
on heterogeneous
systems~\cite{ompss-gpu-fpga}.



Despite the strong potential of the task-based paradigm, 
its effective advantages are far from being well understood when applied to
the non-trivial programs that comprise real-world HPC applications.
This paper contributes to a better assessment of the advantages and limitations of tasks with data dependencies when used to parallelize the important class of particle-mesh applications. 
The case used for our study is a plasma physics kinetic simulation,
based on an electromagnetic particle-in-cell (EM-PIC) method. This method is widely used for modeling many relevant plasma physics scenarios, ranging from high-intensity laser-plasma interaction to astrophysical shocks~\cite{0741-3335-57-11-113001}. 


This paper makes two main contributions. 
As a \textrm{first contribution}, 
we propose different task-based implementations of 
a bare-bones version of the OSIRIS EM-PIC code~\cite{10.1007/3-540-47789-6_36},
called ZPIC. The different versions
explore the task-based paradigm to different extents -- ranging from its most basic use to advanced features such as data dependencies.
The suite of parallel implementations is available 
as open source to the community\footnote{\url{https://github.com/epeec/zpic-epeec}}, and constitutes a useful benchmark to evaluate future advances in task-based programming tools and HPC hardware.

As a \textrm{second contribution}, we experimentally evaluate these different implementations with realistic simulation workloads (namely, Laser Wakefield Accelerator and Collision of Plasma Clouds) on a shared-memory multicore processor. 
Our results show that a fully asynchronous implementation ({\it i.e.}, using only data dependencies for synchronization) is able to achieve near perfect scaling for 48 cores, despite the unbalanced conditions. This impressive result is accomplished while retaining the code simplicity of task-based programming.  

The remainder of this paper is organized as follows.  Section~\ref{sec:backg} 
provides background on  task-based programming models and on EM-PIC methods.
Section \ref{sec:implementation} describes the proposed parallel implementations. 
Section~\ref{sec:evaluation} 
presents our experimental evaluation. 
Section \ref{sec:relwork} surveys related work. Finally, Section~\ref{sec:conclusion}  presents final remarks and perspectives for future work.

\section{Background} 
\label{sec:backg}

\subsection{Shared Memory Programming}
\label{subsec:sm}


Under a shared memory model, a computational system is composed of
processors that share the same memory space.
Shared memory systems support many different programming paradigms 
but developers tend to prefer high-level programming models, such as those based on directives, seeking high coding productivity. In this paper, we focus on the widely-used OpenMP API as well as the OmpSs tasking model. 


OpenMP~\cite{OpenMPSpecification} is a popular application programming interface (API) for expressing parallelism in shared-memory systems.
The directives provided by Open\-MP allow a simple and incremental parallelization approach from sequential code. The two main work sharing directives are \texttt{for} and  \texttt{sections}. The former allows to distribute loop iterations across threads (data parallelism) and the latter allows to define chunks of code that can run concurrently by different threads (functional parallelism). However, a simple use of these directives can easily assign different amounts of work to the different threads, creating a load imbalance that can lead to idle CPU time due to the synchronization of the different
processors~\cite{10.1371/journal.pone.0077742}. In version 3.0, the  \texttt{task} directive was introduced, which allowed a more dynamic assignment of work to the threads. With tasks, the programmer only needs to identify units of independent work,  leaving the decision about when to schedule their execution by an available thread to the runtime system.
To enforce cross-task coordination (for instance, to ensure that a given segment of code only starts executing after a set of preceding tasks have finished), directives such as \texttt{taskwait} are provided.

More recently, OpenMP 4.0 extended the task construct to allow for defining data dependencies among tasks. The  \texttt{in} clause prevents a task from being scheduled before the variables specified in the clause are available. The threads that produce these variables will in turn specify this information with the \texttt{out} clause, meaning that the former thread will only start after all of these have finished. It is also possible to specify an \texttt{inout} dependency. The data dependencies among the tasks define a data-flow graph whose operations are executed asynchronously as the necessary data becomes available. These clauses may also be used to reduce the number of synchronization directives among tasks, since the dependencies may be used to guarantee mutual synchronization implicitly. Data dependencies in OpenMP tasks were introduced after the SMPSs programming model~\cite{smpss}, a precursor of OmpSs~\cite{ompss}. Since then, different task-related improvements have been studied within OmpSs, such as accelerator offloading~\cite{ompss-gpu-fpga}
or task-parallel reductions~\cite{10.1007/978-3-319-11454-5_1}. Several of those are already adopted by the OpenMP standard, while some others are under current active discussion. Besides OmpSs and OpenMP, there are other programming models that support tasking, such as StarPU~\cite{augonnet2011starpu}, 
Cilk~\cite{cilk}, Intel TBB~\cite{intel_tbb}, 
and High Performance ParalleX (HPX)~\cite{HPX}.

Initially, our intention was to focus only on the OpenMP tasking model, however current implementations do not fully support all types of data dependencies. This is the case of {\tt mutexinoutset} ({\tt commutative} in OmpSs), which allows mutual dependencies between tasks, but without a predefined order of execution. For this reason, all task-based implementations of this paper were developed in OmpSs-2, the second generation of the OmpSs programming model.

\subsection{Kinetic Plasma Simulations} 
\label{sec:plasmasimulations}

Electromagnetic particle-in-cell (EM-PIC) codes such as OSIRIS \cite{Fonseca_2013} have found widespread use in modeling the highly nonlinear and kinetic processes that occur in several relevant plasma physics scenarios, ranging from astrophysical settings to high-intensity laser plasma interaction. In an EM-PIC code, the full set of Maxwell equations is solved on a grid using currents and charge densities calculated by weighting discrete particles onto the grid~\cite{1999JPlPh..61..425P}. Each particle is then pushed to a new position and momentum via the self-consistently calculated fields. Therefore, to the extent that quantum mechanical effects may be neglected, an EM-PIC code makes no physics approximations and is ideally suited for studying complex systems with many degrees of freedom. For the analysis in this paper, a simplified version of OSIRIS called ZPIC~\cite{ZPIC} was considered. ZPIC is a purely sequential, bare-bones EM-PIC code implementing exactly the same algorithm as OSIRIS, and maintains all the core features of the latter. Therefore, the ZPIC code is relatively simple yet accurate, allowing an easy exploration of different programming models and parallel platforms.


\begin{figure}[t]
 \centering
 \includegraphics[width=96mm]{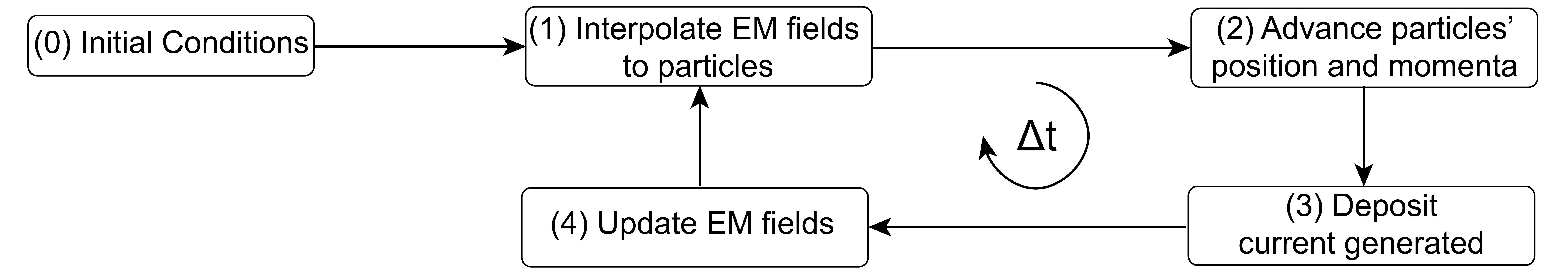}
 \caption{Main stages in an EM-PIC simulation (adapted from~\cite{0741-3335-47-5A-017}).}
 \label{fig:PIC_cycle}
 \squeezeup
\end{figure}

The main simulation loop of EM-PIC methods is usually divided into four stages~\cite{0741-3335-47-5A-017}, as depicted in
Figure~\ref{fig:PIC_cycle}. 
In ZPIC, the field interpolation stage (1) is done using a bi-linear interpolation with the field values from the previous iteration, which are then used for the calculation of the Lorentz force acting on each individual particle. The particle advance stage (2) determines the next position and momenta, by integrating equations of motion using a leapfrog scheme~\cite{0741-3335-47-5A-017} 
and the Boris method~\cite{0741-3335-57-11-113001}, 
making this method second order accurate in time. Using the particle motion calculated in the previous stage, the next stage in ZPIC~(3) determines the electric current density on the grid using an exact charge conserving method~\cite{VILLASENOR1992306}. The code may apply a digital filter on the current density to reduce short wavelength noise. As a last stage (4), using the current density that was just calculated, the code advances the EM fields in time using a finite-difference time-domain technique on a Yee mesh~\cite{0741-3335-57-11-113001,0741-3335-47-5A-017}. 

Also note that the EM-PIC algorithm described here is an implementation of the more general class of particle-mesh algorithms suited for (relativistic) kinetic plasma simulations. In this class of algorithms, the interaction of a large set of particles (bodies) is mediated by fields deposited on a finite mesh, instead of using a direct interaction between the particles. This allows for the algorithm complexity to scale with $\simeq N_p$ (the number of particles) rather than $\simeq N_p^2$ while retaining correct results as long as short range ({\it i.e.}, shorter than the mesh cell size) interactions do not dominate. This large computational gain makes this class of algorithm extremely popular in many fields, and, while this paper focuses on EM-PIC, the results presented here can be readily applied to any other particle-mesh code.

\section{Parallel EM-PIC Implementations} 
\label{sec:implementation}




We propose a diverse set of implementations, each covering a relevant point in a vast design space. We start with a natural parallelization of the original ZPIC code, which does not exploit tasks, and remains close to the original program structure. We then depart to task-based implementations, exploring different tasking features.

\subsection{Parallel For-based Implementation} \label{subsec:typical_impl}

This initial implementation, which we call \texttt{zpic-parallel-for}, is the most incremental approach to the original ZPIC code. 
We use OpenMP's \texttt{for} directive to naturally exploit the inherent data parallelism of the the original loop structure of the (sequential) ZPIC code. 



Recalling the four main stages of the EM-PIC code (Figure \ref{fig:PIC_cycle}), the first three stages are implemented as a single loop that iterates over all the particles in the simulation. Each iteration interpolates the EM fields at the particle position (stage 1), advances the its momentum and position (stage 2) and then deposits the generated current in the grid (stage 3). Considering that the particles in EM-PIC implementation do not interact directly, but rather through the grid, each particle can be advanced independently (stages 1-2). Therefore, the most natural way to parallelize these three stages is to distribute the particles evenly among the threads with an OpenMP \texttt{for} directive ({\it i.e.}, a particle-based decomposition). However, since all threads share the same global buffer with the grid quantities (electric current and EM fields), two or more threads can advance closeby particles and try to deposit their currents in the same cell, causing a data race. A simple solution is to update the electric current atomically. However, according to our experiments, this approach often results in poor performance. Instead, we created per-thread copies of the electric current buffer, so that each thread can deposit the current in its copy without interfering with the others threads. After all particles (from all threads) have advanced in a given iteration, the program combines all the copies into a single buffer using OpenMP's reduction mechanism. Depending on the grid size, the number of time steps and plasma density, this global large-scale reduction can severely limit the scalability of the \texttt{zpic-parallel-for}. 

Once stages 1-3 are complete, the program assigns a range of rows in the global grid to each thread (again, using OpenMP's \texttt{for} directive), whose EM fields are updated in parallel. There are no data races in this stage of the simulation.



In this implementation, there are several global synchronization points to ensure that either the electric current or the EM fields are updated completely ({\it i.e.}, in the entire simulation space) before proceeding to the next stage of the simulation. Therefore, if some threads happen to receive a higher load in a given simulation stage, they will straggle, forcing other threads to linger at the synchronization point.

Since the global reduction and synchronization are hard to avoid in a particle-based decomposition, we must change our parallelization strategy to improve the program scalability and efficiency. We describe this approach next and how it can be complemented by a tasking model.

\subsection{Task-based Implementations} \label{subsec:task_impl}

In the tasking model, we define work units as \texttt{tasks} and rely on the runtime to schedule tasks to the threads rather than resorting to parallel \texttt{for} loops to assign work ({\it e.g.}, particles or grid rows) to threads.
Furthermore, by overdecomposing the problem ({\it i.e.}, creating more concurrent tasks than the available cores), we expect the runtime to mitigate load imbalances by dynamically assigning new tasks to threads that become idle after having completed a previous task.

The success of this strategy depends on the programmer's ability to minimize the synchronization restrictions associated with each spawned task.
Ideally, we would like to minimize scenarios where a task needs to synchronize globally ({\it i.e.}, with all the other concurrent tasks), and replace them with local synchronization ({\it i.e.}, a task needs to coordinate its actions with a few concurrent tasks).

In the case of ZPIC, the limited scalability and the presence of global synchronization points made us abandon the natural particle-based decomposition for stages 1-3. Instead, we adopt a spatial decomposition for the entire simulation loop, similar to many state-of-the-art EM-PIC codes \cite{Fonseca_2013,psc,smilei}. In this approach, the simulation space is split into \emph{regions} alongside the $y$ axis  ({\it e.g.}, a row-wise decomposition). Each region stores both the particles inside it and the fraction of the grid they interact with, allowing both the particle advance and field integration to be performed locally. 

However, implementing this spatial decomposition is more complex. Particles can exit their assigned region and must be transferred to an adjacent region. Each region must also be padded with ghost cells (extra rows of cells at the top and bottom of the assigned region, which are copies of neighbor regions), so that the thread processing the region can access grid quantities outside its boundaries. In both cases, the communication and synchronization of a given task then become limited to the couple of tasks that manage the adjacent regions. (Note that, due to simulation conditions, a particle can only move to a neighbor cell at each time step.) 

Tasks can be synchronized in two ways: with explicit barriers or data dependencies. The latter can be more efficient if the data dependencies are defined in a way that prevents data races and allows for a more efficient task scheduling. Then, the runtime can schedule tasks as soon as their dependencies are satisfied.
Unrelated tasks will be executed asynchronously. 


\begin{figure}[t]
 \centering
 \includegraphics[width=0.9\linewidth]{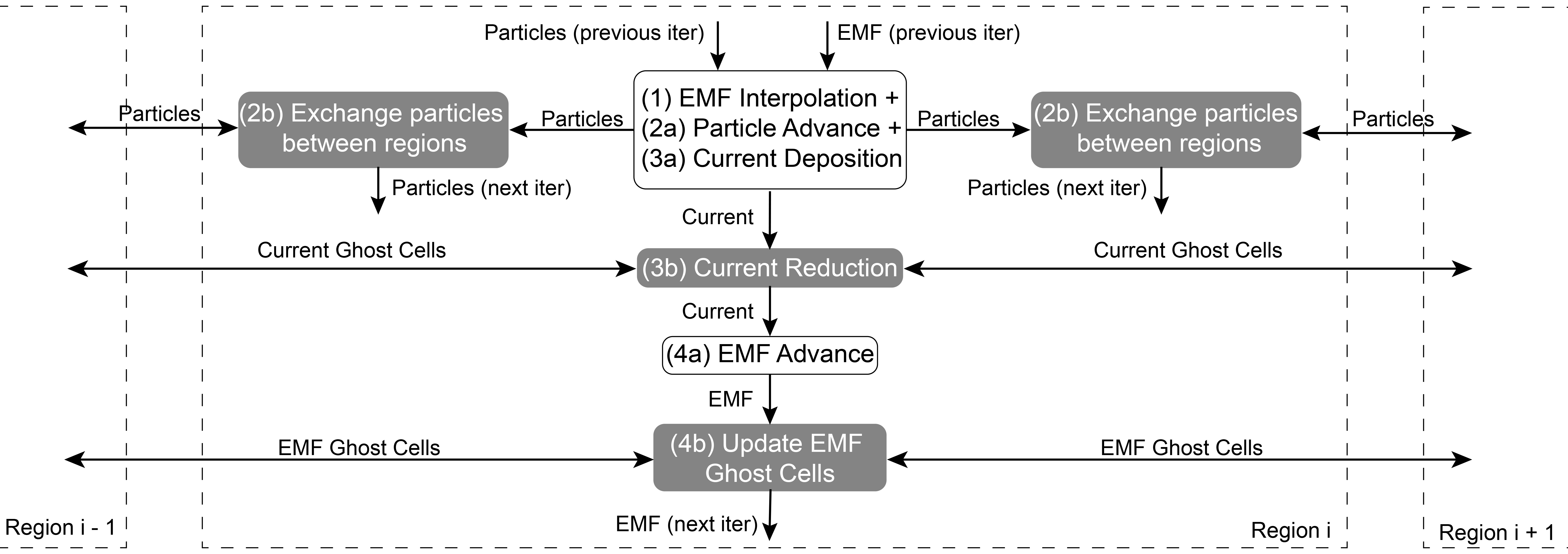}
 \caption{Data flow (arrows) between the main tasks (round boxes) for one region within a single time step. A spatial decomposition requires additional tasks (in gray) for exchanging data between neighbor regions.}
 \label{fig:dataflux}
 \squeezeup
\end{figure}

Figure \ref{fig:dataflux} shows the tasks for a single region and the data flow between them. Each iteration of the simulation loop begins with the particle advance, in which a task advances all particles within a given region (stages 1-3). However, two tasks can advance particles near the boundaries of a neighbor region and deposit the current in the same cell, causing a data race (similarly to Section~\ref{subsec:typical_impl}). One solution is to store the electric current in local buffers and perform a reduction operation to obtain the final current. Differently from Section \ref{subsec:typical_impl}, the reduction is only required for ghost cells and can be executed completely in parallel.

Another solution is to use a global buffer and synchronize access to this buffer through data dependencies.
This buffer synchronization requires an \texttt{inout} clause creating a mutual data dependency between tasks handling adjacent regions. The way the runtime solves this dependency loop is by executing the tasks in order of their creation. In our case, this would create a sequential execution of all tasks. To avoid this, we use the \texttt{commutative} clause of OmpSs, allowing threads to advance particles in adjacent regions in any order, but not at the same time. 


After calculating the final electric current, the next set of tasks advances the electromagnetic fields in each region (stage 4). 
At the end of the simulation loop, each regions updates the values of the EM fields in their ghost cells.

From the same base algorithm and spatial decomposition, we implemented variants of the code to test the different tasking features. Table~\ref{table:task_variants} provides a brief comparison between the different versions. In the \texttt{tasklike} variant, an OpenMP parallel \texttt{for} loop dynamically assigns each region to a thread, one at a time. The thread then executes all the tasks of the associated region. In this case, the tasks from different regions are synchronized through for-loop barriers. All the other variants are implemented in OmpSs. The \texttt{async} suffix indicates that tasks are synchronized exclusively by data dependencies and the program execution is completely asynchronous. In the \texttt{sync} variant, in turn, a barrier at the end of each time step ensures that all the tasks have completed before executing the next time step.

\setlength{\tabcolsep}{0.5em} 
{\renewcommand{\arraystretch}{1.2}
\begin{table}[t]
\scriptsize
\centering
\caption{Features of each task-based implementation.}
\makebox[\linewidth]{
\begin{tabular}{|c|c|c|c|c|}
\hline
Version &
  Synchronization &
  Data Race &
  Asynchronous? \\ \hline
\texttt{zpic-tasklike}       & Barriers          & Reduction   & No   \\ \hline
\texttt{zpic-reduction-sync} &
  \begin{tabular}[c]{@{}c@{}}Data Dependencies \\ (Barrier at the end)\end{tabular} &
  Reduction &
  Partial \\ \hline
\texttt{zpic-commutative-sync} &
  \begin{tabular}[c]{@{}c@{}}Data Dependencies \\ (Barrier at the end)\end{tabular} &
  Commutative &
  Partial \\ \hline
\texttt{zpic-reduction-async}   & Data Dependencies & Reduction   & Full \\ \hline
\texttt{zpic-commutative-async}  & Data Dependencies & Commutative & Full \\ \hline
\end{tabular}
}
\label{table:task_variants}
\squeezeup
\end{table}
}

As stated before, a major problem in a spatial decomposition with fixed regions is load imbalance. In some simulations, the particle movement will cause some regions to have a higher plasma density than the others, even if the initial distribution is uniform. Hence, some tasks will take much longer than others to complete. 
As a solution, 
we overdecompose the simulation space in more regions (thus, more tasks) than the number of available threads. The created tasks are then dynamically distributed to the threads as a way to balance the load among them. The balance granularity is determined by the region size and smaller regions often lead to smoother load distributions. Smaller regions may also result in a better cache usage as the working set of each (smaller) task may now fit in the L1 cache. 
Naturally, the cost of an overdecomposition is the additional communication and synchronization between regions. In a shared-memory environment, 
the communication between regions consists in copying data from one memory position to another, which is a fairly cheap operation as long as the region is reasonably sized.
According to our experiments (Section~\ref{sec:results}), ZPIC performs best when using 2-3x as many regions as the number of cores. 
Both SMILEI~\cite{smilei} and PSC~\cite{psc} uses a similar load balancing technique.

        


        



\subsection{Code Complexity}
Regardless of the target programming model, a spatial decomposition (Section~\ref{subsec:task_impl}) is more complex to implement than a particle-based decomposition (Section~\ref{subsec:typical_impl}). In a spatial decomposition, the program has to split the simulation space into multiple regions (each one with separate buffers), treat each region individually and handle all the communication between them. In contrast in a particle-based decomposition, we directly exploit the inherent data parallelism of the \texttt{for} loops of the original sequential implementation. Both strategies are common in parallel EM-PIC codes.

\begin{listing} [t]
\caption{The electromagnetic fields advance in the tasking (left) and parallel \texttt{for} (right) paradigms.}
\label{code:emf_adv}
\scriptsize
\begin{minipage}{0.5\textwidth}
 \begin{verbatim}
    #pragma oss task \ 
        inout(E[0; size]) \ 
        inout(B[0; size]) \ 
        in(J[0; size])
    void emf_advance(...);
    
    for(i = 0; i < n_regions; i++)
        emf_advance(...);
\end{verbatim}
\end{minipage}
\vline
\begin{minipage}{0.5\textwidth}
\begin{verbatim}
    void emf_advance(...);
    
    #pragma omp parallel for
    for(i = 0; i < n_regions; i++)
        emf_advance(...);
\end{verbatim}
\end{minipage}
\end{listing}

Considering the same decomposition, both tasking and parallel \texttt{for} paradigms have similar code complexity (Listing \ref{code:emf_adv}), with less than 1\% difference in terms of lines of code. However, a \texttt{task} directive may carry additional information in the form of data dependencies. The runtime then uses these dependencies to synchronize and schedule the tasks, in a way that is transparent to the programmer. In contrast, all synchronization points in a parallel \texttt{for} need to be explicitly defined by the programmer.
\section{Evaluation}
\label{sec:evaluation}

The main goal of our evaluation is to study how the proposed parallel implementations of ZPIC scale when 
used to simulate realistic large-scale problems.
By doing that, we assess whether the virtues of the task-based paradigm, especially when complemented with data dependencies, effectively translate to relevant performance gains.



\subsection{Experimental Methodology}
\label{sec:setup}

All programs were compiled with GNU GCC 10.1 (OpenMP v4.5) with the \texttt{-O3} optimization level. For OmpSs programs, we used the OmpSs-2 2020.06 release version with GCC as the back-end compiler.
The  results were obtained on a computational node composed of two Intel Xeon Platinum 8160 CPUs with $24$ physical cores @2.10GHz (total of 48 cores) and 96GB of RAM, running SUSE Linux.


The presented results are the average across five runs. We observed a maximum standard deviation of 1.7\%. The speedup was calculated based on the execution time of the original, sequential implementation. There is only one thread running in each CPU core.

To evaluate the performance and correctness of all parallel implementations, we used three types of simulations, which we summarize next.

\begin{figure}[t]
\centering
\includegraphics[width=0.75\linewidth]{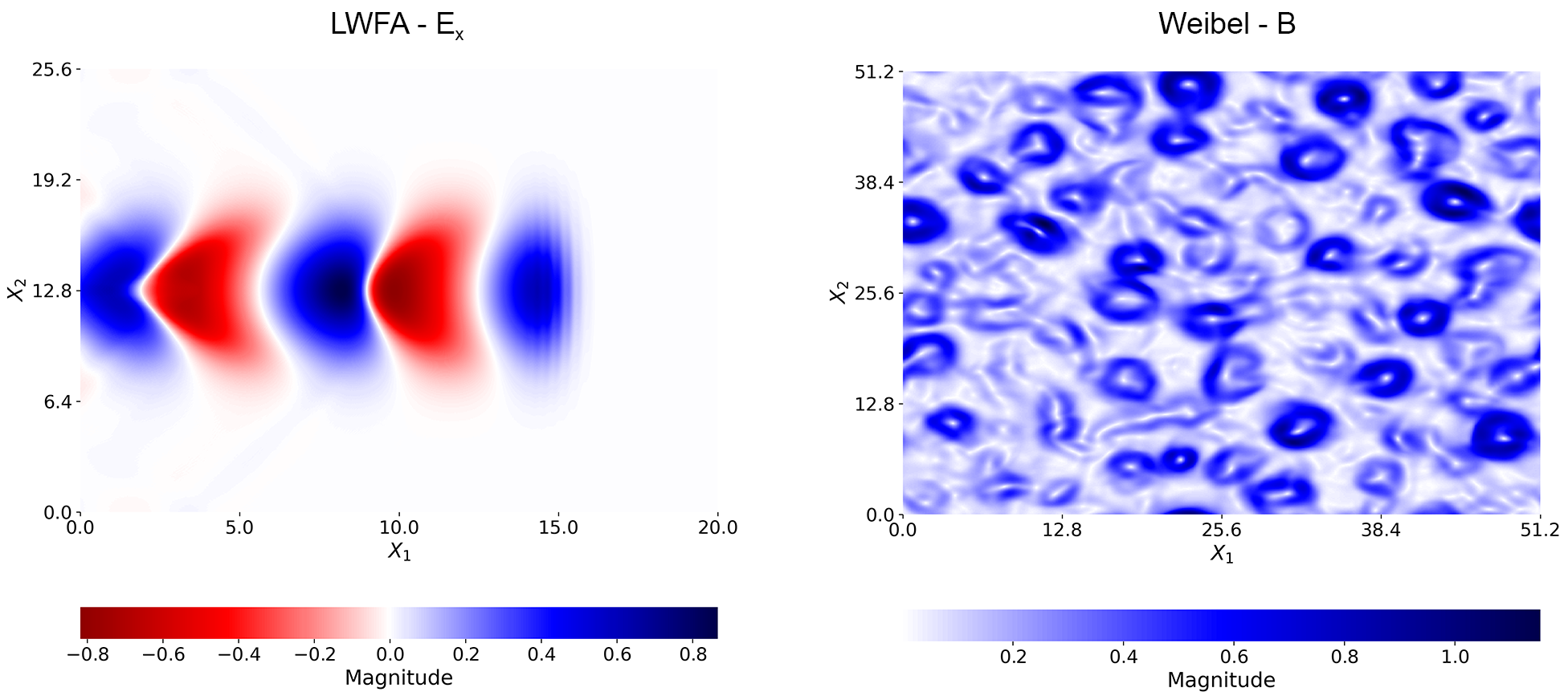}

\includegraphics[width=0.75\linewidth]{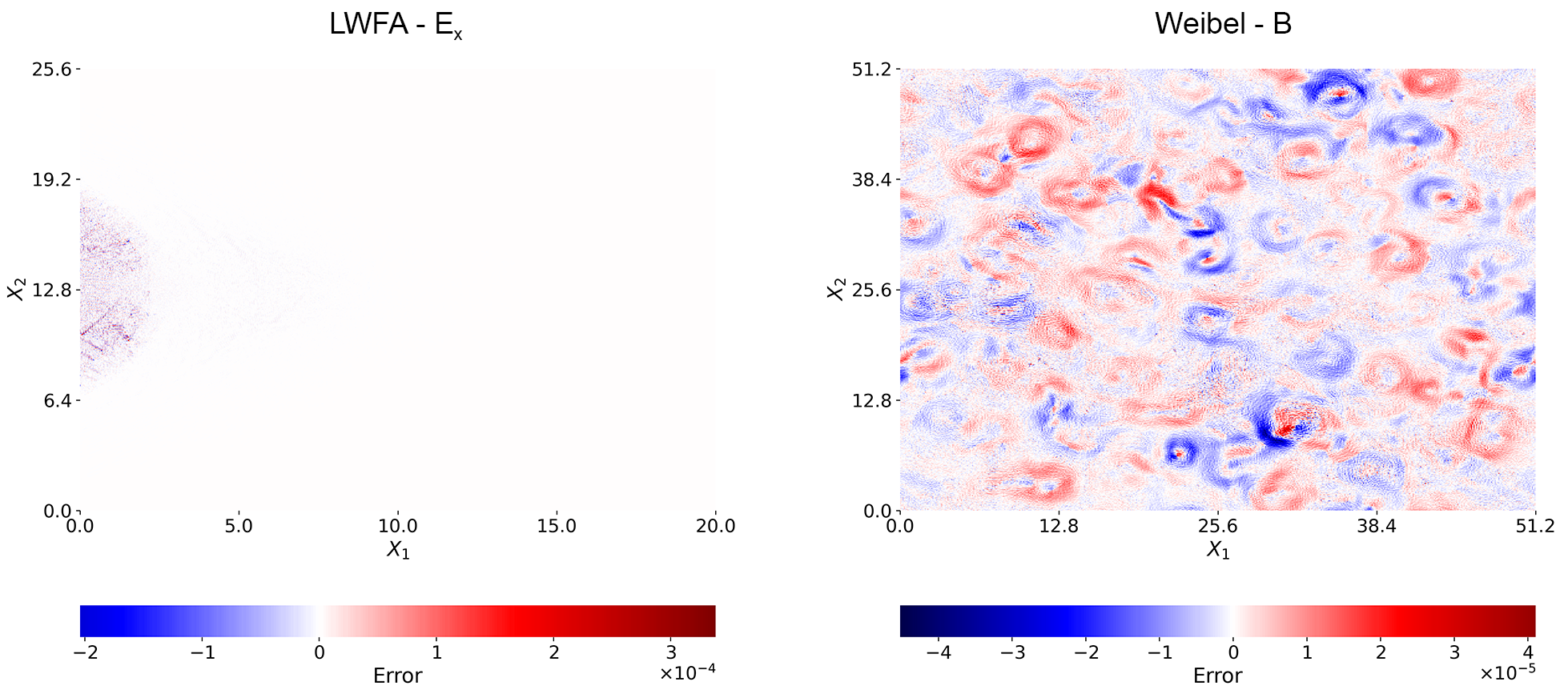}
\caption{(Above) Field report for the last time step in both LWFA and Weibel simulations. (Below) Relative error difference between the results of the \texttt{zpic-reduction-async} implementation and the sequential baseline, for both LWFA and Weibel simulations.} 
  \label{fig:experiment_fields}
\end{figure}

\textbf{Laser Wakefield Accelerator~\cite{PhysRevLett.43.267} (LWFA).} 
In this scenario, a high intensity, short duration laser pulse propagates through an initially uniform plasma. The interaction of the laser with the plasma leads to the formation of a wake trailing the laser pulse, which has an intense longitudinal electric field that can be used to accelerate charged particles, including particles from the background plasma itself (Figure \ref{fig:experiment_fields}, above). In this test case, there is only one plasma species consisting of electrons (the ions are assumed to form an immobile neutralizing background), and the simulation EM fields are initialized with the field values of the laser. The simulation grids were $2000\times 512$ cells, and the particles were initialized with 16 particles per cell. The simulation was run for $4000$ time steps, and used a compensated binomial filter for the current. 
A moving simulation window follows the laser as it propagates through the plasma.

\textbf{Collision of Plasma Clouds~\cite{doi:10.1063/1.1556605} (Weibel).} 
In this case, ZPIC models two plasma clouds moving perpendicular to the simulation plane. One of the clouds is made of electrons and the other is made of positrons. These clouds start with the same initial density and temperature, but move in opposite directions. This system is susceptible to the so called Weibel instability, that leads to the generation of magnetic field and to the filamentation of the plasma clouds (Figure~\ref{fig:experiment_fields}). This test case used a grid size of $512\times 512$ cells, with each of the plasma species using 256 particles per cell, uniformly distributed. The simulation runs for $500$ time-steps. 

\textbf{Uniform Plasmas. } 
We also benchmark our implementation against an isolated, infinite, uniform plasma in two scenarios: a \texttt{cold} plasma, where all particles are initialized at rest, and a \texttt{warm} plasma, where particles are initialized from a thermal distribution with a width $u_{th} = 0.01 \mathrm{c}$. There are no initial flow velocity or EM fields. These scenarios are ideal for peak performance benchmarks, as particle density is expected to remain uniform over the simulation space, and there is limited (\texttt{warm}) to no (\texttt{cold}) particle motion over the simulation. These instances were used exclusively on the weak scaling test.


%

\subsection{Validation of the Parallel Implementations}

To validate the parallel implementations, we compare the last report of the magnetic field map generated by each of the parallel implementations and the original ZPIC, and then calculate the differences between them. The maximum relative error observed is on the order of $10^{-4}$. Figure~\ref{fig:experiment_fields} (below) illustrates this with the \texttt{zpic-reduction-async} implementation.

The discrepancies between the sequential baseline and other implementations are related solely to the electric current deposition algorithm because the order in which the current for each particle is accumulated on the grid changes (when introducing concurrency), leading to different roundoff errors. The maximum error observed for this number of iterations is on par with what is to be expected if we were to randomize the position of the particles in the buffer in the serial implementation and, given that both implementations use exactly the same analytical formalism, one cannot argue that one implementation more accurately models the system than the other. This discrepancy can be significantly reduced by performing the calculations in double precision. However, it should again be noted that our implementation has no effect on the numerical stability of the PIC method and that it has no bearing in the macroscopical physical results. 

\subsection{Results}
\label{sec:results}

\begin{figure}[!tb]
\centering
\includegraphics[width=\linewidth]{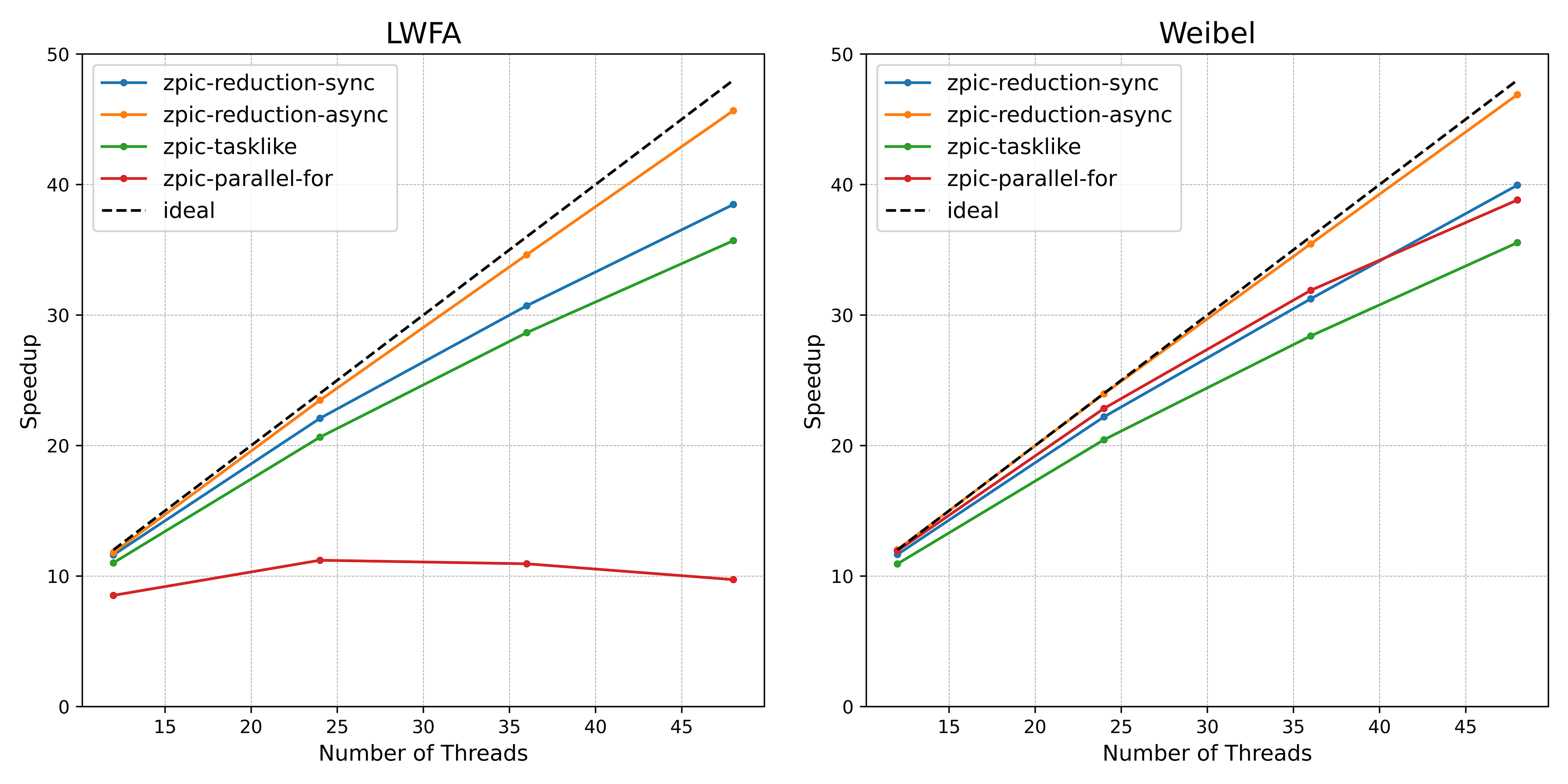}
\caption{Scalability comparison among all reduction-based implementations.} 
 \label{fig:strong_scaling}
 \squeezeup
\end{figure}

As our first experiment, we compare the strong scaling between all reduction-based implementations of ZPIC (Figure \ref{fig:strong_scaling}). In this experiment, we fix the number of regions at 144 (which determines the number of concurrent tasks spawned at any given stage of the simulation) -- which clearly over-decomposes the problem, given that there are only 48 cores available. We evaluate the impact of changing the number of regions later. 

Considering that the Weibel simulation has a high plasma density, 
the cost of \texttt{zpic-parallel-for}’s reduction is amortized over a large number of particles. This diluted cost, combined with the fine-grain parallelism of OpenMP \texttt{for} loops, leads to good performance. The opposite happens in the LWFA simulation: the reduction is very expensive compared to the amount of work assigned to each thread (there are {\raise.17ex\hbox{$\scriptstyle\sim$}}16x less particles per thread than in the Weibel simulation). Since the cost of this operation increases with the number of threads (and corresponding copies), \texttt{zpic-parallel-for} scales very poorly in the LWFA simulation, even having negative scaling over 24 cores.

After changing from a particle-based to a spatial decomposition, the performance of the program is no longer dependent on the cost of the reduction operation, since this operation is restricted to the ghost cells and can be performed completely in parallel. As a result, there is only a 1-2x speedup difference between the LWFA and the Weibel simulation for any \texttt{reduction} variant and \texttt{zpic-tasklike}. At the same time, the program now explores more coarse-grain parallelism (each thread processes a set of regions instead of iterations in a loop) compared to \texttt{zpic-parallel-for}. 

The performance of the task-based implementations depends on the synchronization method. Since \texttt{zpic-tasklike} relies on frequent and costly global barriers, this version has the worst performance among them. Replacing these barriers with data dependencies not only improves the runtime's load balancing capabilities but also lowers the synchronization costs. Combined with an overdecomposition, \texttt{zpic-reduction-sync} is able to match or surpass \texttt{zpic-parallel-for} even in its best case, whereas a fully asynchronous implementation (\texttt{async} variant) achieves near-perfect scaling for 48 cores with a maximum speedup of 46.89x for Weibel and 45.67x for LWFA. The \texttt{commutative}-based versions are discussed in the next experiment.

\begin{figure}[!tb]
\centering
\includegraphics[width=\linewidth]{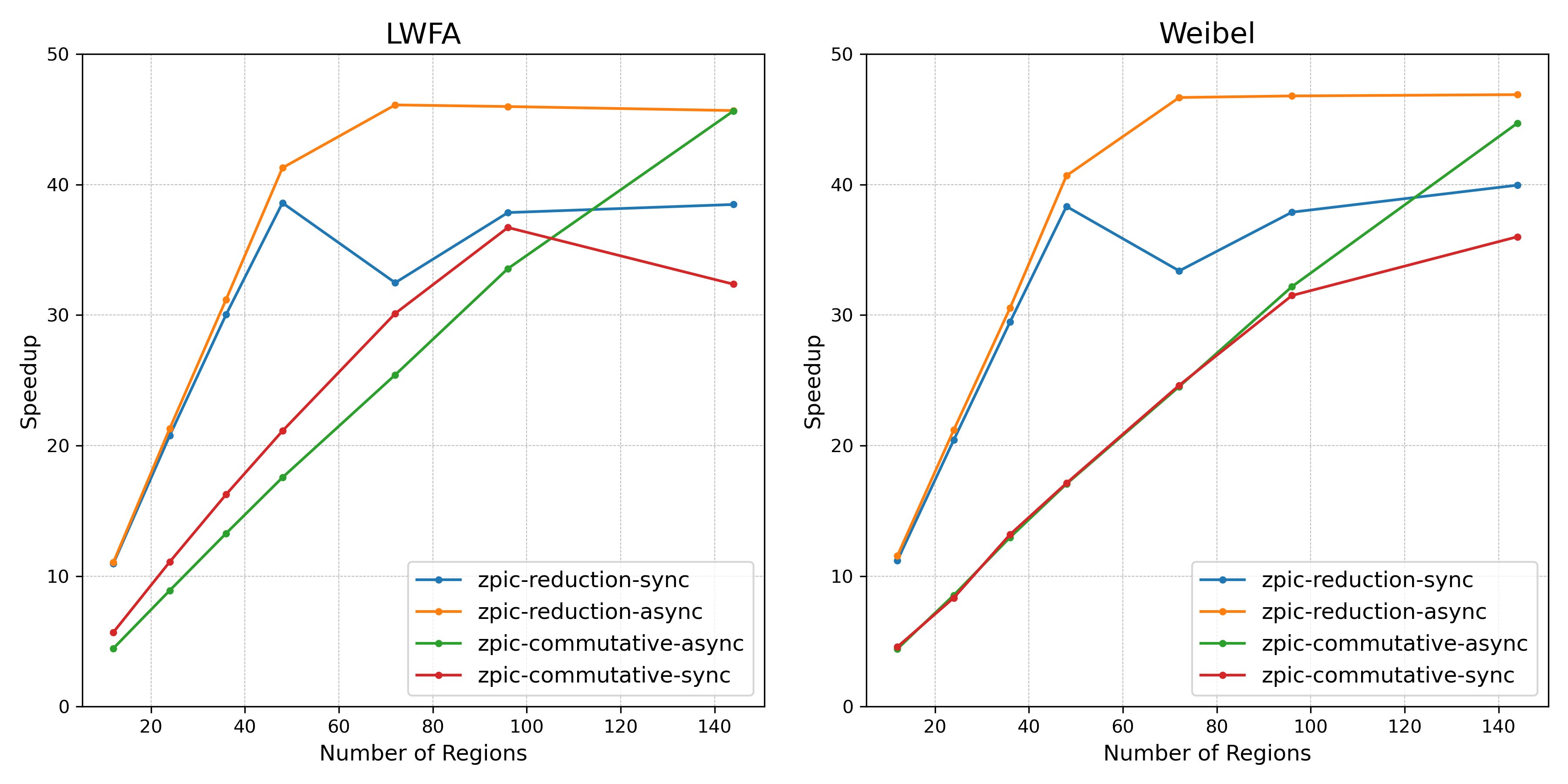}
\caption{Performance comparison between task-based implementation of ZPIC for different number of regions.} 
 \label{fig:region_test}
 \squeezeup
\end{figure}

In the next experiment, we analyze the performance impact of the number of regions for different task-based implementations (Figure \ref{fig:region_test}). We also compare the difference between the \texttt{reduction} and \texttt{commutative} solutions. 

In the \texttt{reduction}-based implementation, all regions can be processed in parallel, since they have separate, local buffers. In contrast, the \texttt{commutative} clause prevents two or more tasks from advancing the particles in adjacent regions, avoiding race conditions during the current deposition. As the regions are processed in an interleaved manner, the \texttt{commutative} variant only has half of the task throughput of the reduction-based implementation. Due to lower throughput and the extra restrictions, the performance of the \texttt{commutative}-based implementation is usually worse than the \texttt{reduction} equivalent, except when employing a high number of regions ({\it e.g.}, 144 regions). 

With less than 48 regions (or 96 regions for the \texttt{commutative} variant), the program is unable to produce enough tasks to fully utilize the CPU. Even with a single concurrent task per thread, the uneven distribution of plasma across the simulation space causes some threads to be idle while waiting for other threads to finish its assigned task. With an overdecomposition ({\it i.e.}, more than one region per core), the program is able to generate enough concurrent tasks to maintain a high CPU occupancy. This happens not only because the program is able to maintain a proper load balance by dynamically distributing the tasks, but also because it can constantly generate new tasks that can be fed to the thread pool. Inserting a barrier at the end of the iteration (\texttt{sync} variant) interrupts the generation of new tasks, causing some threads to be waiting for the last few tasks of the iteration to finish before advancing to the next iteration. As a result, any \texttt{sync} version has negligible performance gains with an overdecomposition. 

In both \texttt{sync} versions, performance drops when the number of concurrent tasks is not evenly divisible by the number of cores ({\it e.g.}, at 72 concurrent tasks). In this case, the tasks of the current iteration will be unevenly distributed among threads, leading to a load imbalance. Without the global barrier, the runtime can schedule tasks from the next iteration to fill the load difference. The \texttt{commutative} clause imposes additional restrictions to the task scheduling, preventing the program from properly balancing the load across threads. 

\begin{figure}[!tb]
\centering
\includegraphics[width=0.8\linewidth]{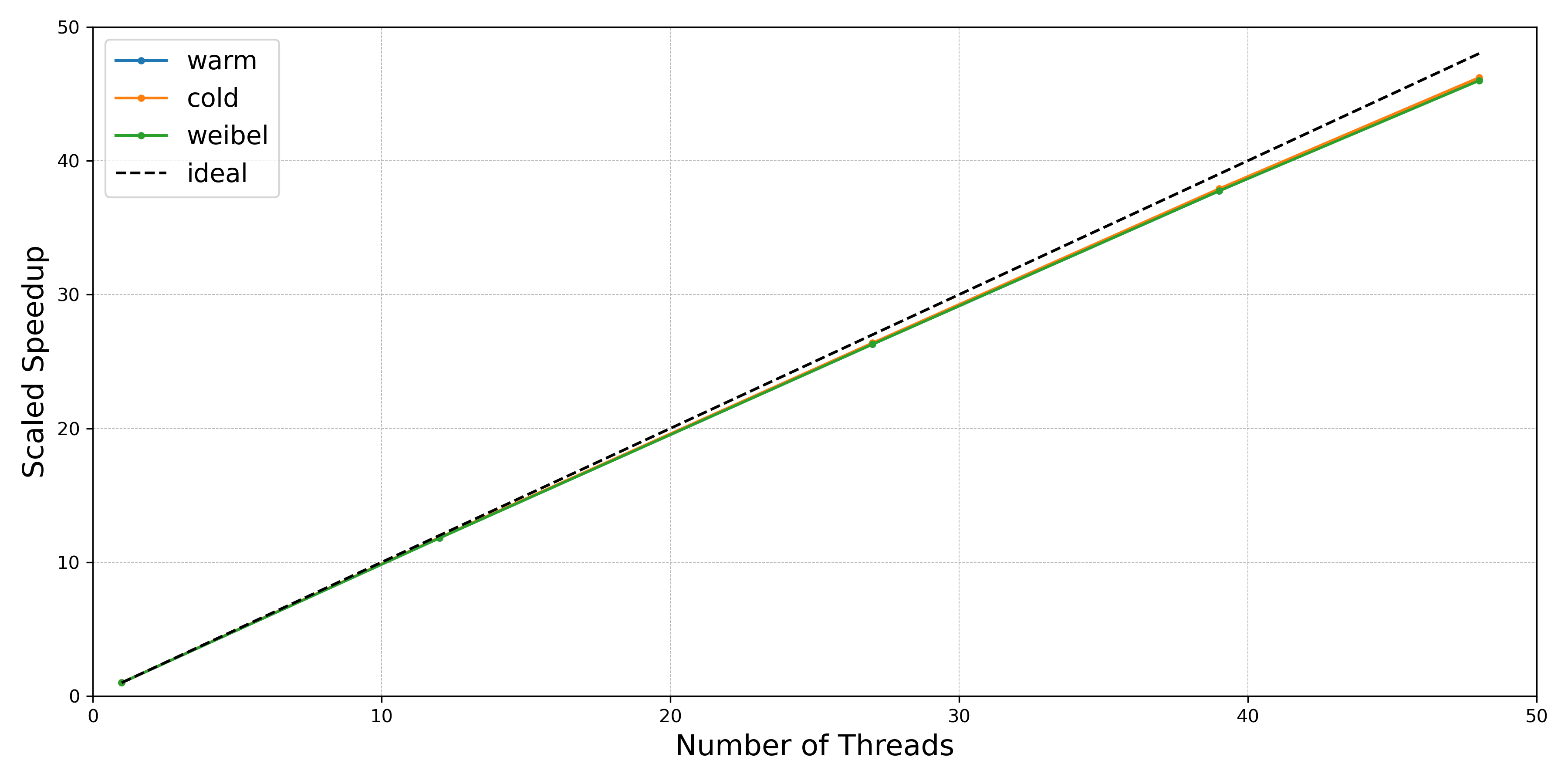}
\caption{Weak scaling for the best performing implementation of ZPIC.} 
 \label{fig:weak_scaling}
 \squeezeup
\end{figure}

Finally, we present in Figure~\ref{fig:weak_scaling} the weak scaling results for the best performing implementation (\texttt{zpic-reduction-async}). For this experiment, we used the Weibel simulation as well as two theoretical plasmas (\texttt{cold} and \texttt{warm}). Using a fixed number of iterations ($500$) and particles per cell ($16\times16$), we vary the grid size to scale the problem with the number of threads. The LWFA simulation cannot be used in this experiment, since the computational load is mostly concentrated on the small region of the simulation space affected by the laser.

The program performs equally well in all simulations, attaining an efficiency greater than 95\% in all experiments. Both \texttt{cold} and \texttt{warm} plasmas have perfect load distribution, as uniform plasma density remains unchanged throughout the entire simulation. In these theoretical plasmas, the particles remain (almost) static, resulting in very little communication between regions. The \texttt{weibel} simulation is completely different: not only can the plasma density vary from one region to another due to the plasma filamentation, but also the particles are rapidly moving throughout the simulation space. Even in these conditions, the program can still maintain an excellent load balance among the threads due to the overdecomposition and dynamic task distribution.

\section{Related Work}
\label{sec:relwork}


Tasking has been studied in the context of specific application domains, such as linear algebra~\cite{aliaga}, human brain simulation~\cite{hbp}, graph analytics~\cite{openmp_graph}, 
adaptive mesh refinement \cite{10.1145/3225058.3225085,10.1007/978-3-030-28596-8_15}, among others~\cite{openmp_others,chasapis_parsecss_2015}. 
However, only a few papers \cite{10.1145/3225058.3225085,10.1007/978-3-030-28596-8_15} focus on the tasking features that are available since OpenMP 4.0.

Regarding particle-mesh algorithms, which are the focus of this paper, tasking is only scantly studied. Akhmetova et al. \cite{akhmetova_performance_2017} used tasking to improve an MPI + OpenMP hybrid EM-PIC code. 
They introduced the \texttt{task} directive (without data dependencies) in existing \texttt{for} loops of the particle solver. 
In contrast, our paper studies the tasking model when applied to the entire EM-PIC algorithm, using data dependencies for synchronization purposes.  
Koniges et al. \cite{cray_tasking} propose the use of OpenMP tasking to hide communication in Gyrokinetic Toroidal Simulation (GTS) code. They use OpenMP 3.0, which does not support data dependencies. Anderson et al.~\cite{jarvis_performance_2014} ported the 3D Gyrokinetic Toroidal Code (GTC) \cite{ethier_gyrokinetic_2005} to HPX \cite{HPX}, exploiting the support of the latter for tasking with data dependencies. In their task-based version, they overdecompose the problem as a way to overlap the communication and balance the load across the CPU cores within a single node.

\section{Conclusions}
\label{sec:conclusion}

We developed and analyzed a set of task-based implementations of an EM-PIC simulator as a way to contribute to a better understanding of the benefits and limitations of tasking models when applied to the broad class of particle-mesh codes. 
%
Our results confirm that tasking, when used with recent data dependencies features, enables the runtime to dynamically schedule highly asynchronous tasks, attaining near ideal scalability even with very irregular workloads.
This impressive result is achieved 
while retaining the simplicity of the tasking model, thus providing the programmer with high coding productivity. 


In the future, we plan to investigate tasking in  distributed environments by extending our task-based implementation to support either message passing or partitioned global address spaces. 
In that context, not only can tasks provide a natural way to hide the inter-node communication, but they can also provide a new perspective on the interaction between the two models. Currently, we are extending our task-based implementation to support hardware accelerators, such as GPUs and FPGAs.

\section{Acknowledgements}
This work was partially supported by Fundação Ciência e Tecnologia (FCT) under grant UIDB /50021/2020 and 
by the EPEEC project, which has received funding from the European Union’s Horizon 2020 research and innovation programme under grant agreement No 801051.

\bibliographystyle{splncs04}
\bibliography{bibliography} 

\end{document}